\documentstyle[12pt]{article}

\newcommand{\be}{\begin{equation}}
\newcommand{\ee}{\end{equation}}
\newcommand{\bi}[1]{\vspace{-3mm} \bibitem{#1}}
\begin{document}


\begin{center}
{\Large \bf Vasily E. Tarasov }
\footnote{\normalsize Theoretical High Energy Physics Department,
Skobeltsyn Institute of Nuclear Physics, \\
Moscow State University, 119899 Moscow, RUSSIA\\
{\large E-mail: tarasov@theory.sinp.msu.ru}
}
\end{center}

\vskip 11 mm

\centerline{\Large \bf Quantization of Non-Hamiltonian Systems }
\vskip 11 mm


\centerline{\large \ Preprint of Skobeltsyn Institute of Nuclear Physics\ }
\vskip 2mm
\centerline{\large \ Moscow State University \ }
\vskip 2mm
\centerline{\large \ SINP MSU  \ \ 2000-33/637  \ }

\vskip 7 mm
{\small In this paper a generalization of Weyl quantization which maps
a dynamical operator in a function space to a dynamical superoperator
in an operator space is suggested. Quantization of dynamical
operator, which cannot be represented as Poisson bracket
with some function, is considered. The usual Weyl quantization
of observables can be derived as a specific case of suggested
quantization if dynamical operator is an operator of multiplication
on a function. This approach allows to define consistent Weyl
quantization of non-Hamiltonian and dissipative systems.
Examples of the harmonic oscillator with friction and a system which
evolves by Fokker-Planck-type equation are considered.
}

\newpage

\section{Introduction}


Canonical quantization of classical observables and states defines a map
of real functions into self-adjoint operators \cite{Ber1}.
A classical observable is described by some real function
$A(q,p)$ from a functional space ${\cal M}$.
Quantization of this function leads to self-adjoint operator
$\hat A(\hat q,\hat p)$ from some operator space $\hat {\cal M}$.
It is known that states can be  considered as a special observable.
Classical state can be described by non-negative-normed
real function $\rho(q,p)$ called density distribution function.
Quantization of a function $\rho(q,p)$ leads to non-negative
self-adjoint operator $\hat \rho(\hat q, \hat p)$ of trace class
called matrix density operator.

Time evolution of an observable $A_t(q,p)$ and a state $\rho_t(q,p)$ in
classical mechanics are described by differential
equations on a function space ${\cal M}$:
\[ \frac{d}{dt}A_t(q,p)={\cal L}_t A_t(q,p) \ , \quad
\frac{d}{dt} \rho_t(q,p)=\Lambda_t \rho_t(q,p) \ . \]
The operators ${\cal L}_t$ and  $\Lambda_t$ act
on the elements of function space ${\cal M}$.
These operators are infinitesimal generators of dynamical semigroups
and are called dynamical operators. The first equation describes
evolution of an observable in the Hamilton picture, and the second
equation describes evolution of a state in the Liouville picture.

Dynamics of an observable and state in quantum mechanics
are described by differential equations on an operator space $\hat{\cal M}$:
\[ \frac{d}{dt} \hat A_t( \hat q,\hat p)=
\hat {\cal L}_t \hat A_t(\hat q, \hat p) \ ,
\quad \frac{d}{dt} \hat \rho_t=\hat \Lambda_t \hat \rho_t \ . \]
Here $\hat {\cal L}_t$ and $\hat \Lambda_t$ are superoperators
(operators act on operators). These superoperators are infinitesimal
generators of quantum dynamical semigroups \cite{Koss1,Koss2,Lind1}.
The first equation describes dynamics in Heisenberg picture,
and the second - in Schroedinger picture.

It is easy to see that quantization of the dynamical operators
${\cal L}_t$ and $\Lambda_t$ must lead to dynamical
superoperators $\hat {\cal L}_t$ and $\hat \Lambda_t$. Therefore,
generalization of canonical quantization for general classical
non-Hamiltonian systems must map operators into superoperators.

Usually the quantization is applied to classical systems
with the dynamical operator ${\cal L} A(q,p)=\{ A(q,p), H(q,p)\}$.
Here the function $H(q,p)$ is an observable which characterizes
dynamics and the function $H(q,p)$ is called the Hamilton function.
Quantization of a dynamical operator which can be represented as
Poisson bracket with the Hamilton function is defined by usual
canonical quantization. Canonical quantization of real functions
$A(q,p)$ and $H(q,p)$ leads to self-adjoint operators
$\hat A(\hat q,\hat p)$ and $\hat H(\hat q,\hat p)$.
Quantization of the Poisson bracket $\{A(q,p),H(q,p)\}$ usually defines
as commutator $(i/\hbar)[\hat H(\hat q,\hat p), \hat A(\hat q,\hat p)]$.
Therefore quantization of these dynamical operators
can be uniquely defined by usual canonical quantization.

Quantization of classical non-Hamiltonian systems is not defined by
usual canonical quantization. It is necessary to consider some
generalization of canonical quantization.
These generalized procedure must define a map
of operator into superoperator.

In this paper a Weyl quantization of classical non-Hamiltonian systems
is considered. The generalized Weyl quantization, which maps a
(linear differential, pseudodifferential) operator on a function space
into a superoperator on an operator space, is suggested.
An analysis of generalized quantization is performed for operator which
cannot be represented as the Poisson bracket with some function.


\section{Weyl Quantization}


In this section the usual method of quantization is considered
\cite{Ber1,kn1,kn2}.
To  simplify formulas let $x^k$, where $k=1,...,2n$,
denote usual coordinates $q^j$ and momentums $p^j$
by $x^{2j-1}=q^j$ and $x^{2j}=p^j$, where $j=1,...,n$.
The basis of the space ${\cal M}$ of square-integrable
functions $A(x)$ is defined by functions
\be
\label{f1}
W(a,x)=exp \ iax \ , \quad ax=\sum^{2n}_{k=1} a_k x^k \ .
\ee
Quantization transforms $x^k$ to operators $\hat x^k$.
Weyl quantization of the basis functions (\ref{f1}) leads to
the Weyl operators
\be
\label{f2} \hat W(a,\hat x)=exp \ ia \hat x \ , \quad
a \hat x=\sum^{2n}_{k=1} a_k
\hat x^k \ .
\ee
The Weyl operators form a basis \cite{kn2,BJ}
of the operator space $\hat {\cal M}$.
Classical observable, characterized by the function $A(x)$,
can be represented in the form
\be
\label{f3}
A(x)=\frac{1}{(2\pi)^n} \int \tilde A(a) W(a,x) da \ , \quad
da=da_1...da_{2n} \ ,
\ee
where $\tilde A(a)$ is the Fourier image of the function $A(x)$.
Quantum observable $\hat A(\hat x)$ which corresponds to $A(x)$
can be defined by formula
\be
\label{f4} \hat A(\hat x)=\frac{1}{(2\pi)^n}
\int \tilde A(a) \hat W(a,\hat x) da \ , \quad da=da_1...da_{2n} \ .
\ee
This formula can be considered as an operator expansion for
$\hat A(\hat x)$ in the operator basis (\ref{f2}).
The direct and inverse Fourier transformations allow
to write the formula for the operator $\hat A(\hat x)$ as
\be
\label{f4'}
\hat A(\hat x)=\frac{1}{(2\pi)^{2n}} \int A(x)
\hat W(a, \hat x -x \hat I) da dx\ .
\ee
The function $A(x)$ is called the Weyl symbol of the operator
$\hat A (\hat x)$. Quantization defined by (\ref{f4'}) is called
the Weyl quantization. Another basis operator leads to different
quantization scheme \cite{BJ,kn1,kn2}.

Lie algebra, Jordan algebra and $C^{*}$-algebra are usually
considered on the spaces ${\cal M}$ and $\hat {\cal M}$.

Lie algebra $L({\cal M})$ on the set ${\cal M}$ is defined by
Poisson brackets
\be
\label{f5}
g_{Lie}(A,B) \equiv \{A(x),B(x)\}=\Psi^{km} \frac{\partial
A(x)}{\partial x^k} \frac{\partial B(x)}{\partial x^m}  \ ,
\ee
where $\Psi^{km}$ is a matrix which is inverse of matrix $\omega_{km}$,
that is $\omega_{km} \Psi^{mk'}=\delta^{k'}_{k}$.
All elements of the matrix $\omega_{km}$ are equal to zero, besides elements
$\omega_{2j-1 \ 2j}=1$ and $\omega_{2j \ 2j-1}=-1$, where $j=1,...,n$.
Quantization of the Poisson bracket (\ref{f5}) usually defines as
commutator
\be
\label{f6}
\hat g_{Lie}(\hat A, \hat B) \equiv \frac{1}{i\hbar} [ \hat A(\hat x),
\hat B( \hat x)]= \frac{1}{i\hbar} \Bigl( \hat A(\hat x) \hat B(\hat x)
- \hat B( \hat x) \hat A(\hat x) \Bigr) \ ,
\ee
The commutator defines Lie algebra $\hat L(\hat {\cal M})$
on the set $\hat {\cal M}$.
Leibnitz rule is satisfied for the Poisson brackets.
As a result, the Poisson brackets are defined by basis Poisson brackets
for $x^k$:
\[ \{x^k,x^m\}=\Psi^{km} \ . \]
Quantization of these relations leads to the canonical commutation relations
\be
\label{ccr}
[\hat x^k, \hat x^m]=i\hbar \Psi^{km} \hat I \ .
\ee
These relations can be written for operators $\hat x^{2j-1}=\hat q^j$
and $\hat x^{2j}=\hat p^j$, in the form
\be \label{ccr2}
[\hat q^j, \hat q^{j'}]=0 \ , \quad [\hat p^j, \hat p^{j'}]=0 \ ,
\quad [\hat q^j, \hat p^{j'}]=i\hbar \delta_{jj'} \hat I \ .
\ee
These relations define $(2n+1)$-parametric Lie algebra
$\hat L(\hat {\cal M})$, called Heisenberg algebra.

Jordan algebra $J({\cal M})$ for the set ${\cal M}$ is defined by the
following multiplication
\[ g_{Jord}(A,B)=A(x) \circ B(x)=A(x)B(x) \ . \]
This multiplication coincides with the usual associative multiplication of
functions.
Weyl quantization of the Jordan algebra $J({\cal M})$ leads to the special
operator Jordan algebra $\hat J(\hat {\cal M})$ with multiplication
\[ \hat g_{Jord}(\hat A,\hat B)=[\hat A, \hat B]_{+}=\hat A \circ \hat B=
\frac{1}{4}[(\hat A+\hat B)^2-(\hat A-\hat B)^2 ] \ . \]
Jordan algebra for classical observables is associative algebra, that is
all associators are equal to zero
\[ (A\circ B) \circ C-A \circ (B \circ C)=0 \ . \]
In general case Jordan algebra associator for quantum observables
is not equal to zero
\[ (\hat A \circ \hat B) \circ \hat C- \hat A \circ (\hat B \circ \hat C)=
\frac{\hbar^2}{4} \hat g_{Lie}(\hat B, \hat g_{Lie}(\hat A, \hat C)) \ . \]


The commutation relation (\ref{ccr}) for the operators $x^k$ leads to
the $2n$-parametric Weyl algebra of the operators $\hat W(a,\hat x)$:
\be
\label{W2}
\hat W(a,\hat x) \hat W(b,\hat x)=\hat W(a+b,\hat x)
exp\{ -\frac{i \hbar}{2} a_k \Psi^{km} b_m \} \ ,
\ee
\be
\label{W3}
\hat W^{*}(a,\hat x)=\hat W(-a,\hat x) \ , \quad
\hat W^{*}(a,\hat x)\hat W(a,\hat x)=\hat I \ .
\ee
The Weyl algebra is involute normed algebra.
The involution corresponds to conjugation (\ref{W3}).
The operator norm defines the norm of Weyl algebra.


\section{Quantization of Hamiltonian Dynamical Operator}


Let us consider quantization of a classical dynamical operator
defined by Hamilton function.
Usually the quantization procedure is applied to classical systems
with dynamical operator
\be
\label{f6'}
{\cal L}=-\{H(x), \ . \ \}=-\Psi^{km} \partial_k H(x) \partial_m  \ ,
\ee
where $\partial_{k}=\partial / \partial x^{k}$.
Here $H(x)$ is an observable which defines dynamics of a classical system.
The observable $H(x)$ is called the Hamilton function. Then
\[ {\cal L} A(x)=\{ A(x), H(x) \} \ . \]
If the dynamical operator has this form, then classical system
is called Hamiltonian system.

Weyl quantization (\ref{f4'}) of the functions $A(x)$ and $H(x)$ lead to
operators  $\hat A(\hat x)$ and $\hat H(\hat x)$. Quantization of
Poisson bracket $\{A(x),H(x)\}$ usually defines as the commutator
$(i/\hbar)[\hat H(\hat x), \hat A(\hat x)]$.
Therefore quantization of dynamical operator (\ref{f6'}) leads to
superoperator
\be
\label{f7}
\hat {\cal L}=\frac{i}{\hbar} [\hat H( \hat x), \ . \ ]=
\frac{i}{\hbar}(\hat H^l( \hat x)-\hat H^r( \hat x) )
\ee
Here $\hat H^l ( \hat x)$ and $\hat H^r(\hat x)$ are left and right
superoperators which correspond to Hamilton operator $\hat H(\hat x)$.
These superoperators are defined by formulas \cite{kn1,kn2}:
\[ \hat H^l \hat A= \hat H \hat A \ , \quad
\hat H^r \hat A= \hat A \hat H \ .  \]
Then
\be \label{f8}
\hat {\cal L} \hat A (\hat x) =
\frac{i}{\hbar} [\hat H( \hat x), \hat A(\hat x) ]=
\frac{i}{\hbar}(\hat H( \hat x)\hat A(\hat x)-
\hat A(\hat x) \hat H( \hat x) )  \ .
\ee

Quantization of dynamical operator, which can be represented as Poisson
bracket with a function, can be defined by usual quantization.
Therefore quantization of Hamiltonian systems is completely
defined by the usual method of quantization.


\section{General Dynamical Operator}


Let us consider the time evolution of classical observable $A_t(x)$,
described by the general differential equation
\[ \frac{d}{dt} A_t(x)={\cal L}(x,\partial_x) A_t(x) \ . \]
Here ${\cal L}(x,\partial_x)$ is an operator on the function space
${\cal M}$, and $\partial_{x}$ is a partial derivative with respect
to $x$. Let us consider operator which cannot be expressed
in the form ${\cal L}(x,\partial_x)A(x)=\{A(x),H(x)\}$
with a function $H(x)$. We would like to generalize the quantization
procedure from the dynamical operators (\ref{f6'}) to general
operators ${\cal L}={\cal L}(x,\partial_x)$.

Let us define the basis operators which generate the dynamical
operator ${\cal L}(x,\partial_x)$.
For simplicity, we assume that operator ${\cal  L}$ is a bounded operator.
Operator $Q^k$ is an operator of multiplication on $x^k$ and
operator $P^k$ is self-adjoint differential operator
with respect to $x^k$, that is $-i\partial / \partial x^k$.
These basis operators obey the conditions:\\
1. $Q^k1=x^k$; $P^k1=0$.\\
2.$(Q^k)^*=Q^k$; $(P^k)^*=P^k$. \\
3. $\overline{Q^{k}A(x)}=Q^k\overline{A(x)}$;
   $\overline{P^{k}A(x)}=-P^k\overline{A(x)}$.\\
4. $[Q^k,P^m]=i\delta_{km}$; $[Q^k,Q^m]=0$; $[P^k,P^m]=0$;
   $[1,Q^m]=0$;  $[1,P^m]=0$.\\
Conjugation operation $*$ is defined with respect to scalar product
\[ <A(x)|B(x)>=\int \overline{A(x)} B(x) dx \ . \]
Commutation relations for the operators $P^{k}$ and $Q^{k}$
define $(4n+1)$-parametric Lie algebra. These relations are analogous
to canonical commutation relations (\ref{ccr2}) for
$\hat q^{j}$ and $\hat p^{j}$ with double numbers of degrees of freedom.

Operators $Q^k$ and $P^k$ allow to introduce operator basis
\be
\label{V1}
V(a,b,Q,P)=exp\{ i(aQ+bP)\} \ ,
\ee
for the linear space ${\cal A}({\cal M})$ of dynamical operators.
The basis operators $V(a,b,Q,P)$ are analogous to the
Weyl operators basis (\ref{f2}).
Note that basis functions (\ref{f1}) can be derived from the
operators (\ref{V1}) by the formula $W(a,x)=V(a,0,Q,P)1$.

The commutation relation for the operators $Q^k$ and $P^k$ leads to
the analog of the Weyl algebra:
\be \label{V2}
V(a_1,b_1,Q,P) V(a_2,b_2,Q,P)=V(a_1+a_2,b_1+b_2,Q,P)
exp\{ -\frac{i}{2}[a_1b_2-a_2b_1] \} \ ,
\ee
\be
\label{V3}
V^{*}(a,b,Q,P)=V(-a,-b,Q,P) \ , \quad V^{*}(a,b,Q,P)V(a,b,Q,P)=1 \ .
\ee
This Weyl algebra is involute normed algebra.
The involution corresponds to conjugation (\ref{V3}).
An operator norm defines the norm of Weyl algebra.

The algebra ${\cal A}({\cal M})$ of bounded dynamical operators
can be defined as $C^{*}$-algebra.
It contains all operators $V(a,b,Q,P)$ and is closed for linear
combinations of $V(a,b,Q,P)$ in operator norm topology.
A dynamical operator can be written as an operator function
in the symmetric form
\be
\label{f9}
{\cal L}(Q,P)=\frac{1}{(2\pi)^{2n}} \int \tilde L(a,b) e^{i(aQ+bP)} dadb \ .
\ee
The function $\tilde L(a,b)$ is square-integrable
function of real variables $a$ and $b$. The function $L(a,b)$ is
Fourier image of the operator symbol for ${\cal L}={\cal L}(x,\partial_x)$.
The set of bounded operators ${\cal L}(Q,P)$ and their uniformly limits
form the algebra ${\cal A}({\cal M})$ of dynamical operators.
We can use the basis operators
\[ V_{QP}(a,b,Q,P)=V(a,0,Q,P)V(0,b,Q,P)=e^{iaQ}e^{ibP} \ . \]
which are analogous to Kirkwood basis \cite{BJ}. This operator
basis associate with the standard ordering of operators $Q$ and
$P$. The operators $V_{QP}$ allow to write a dynamical operator in
the $QP$-form
\be
\label{f91}
{\cal L}(Q,P)=\frac{1}{(2\pi)^{2n}} \int \tilde L(a,b) e^{iaQ} e^{ibP} dadb \ .
\ee
This form is suitable for the classical systems.


\section{Quantization of Basis Operators}


To define the superoperator $\hat {\cal L}$
which corresponds to operator ${\cal L}$ we need to describe
Weyl quantization of the basis operators $Q^k$ and $P^k$.
Let us require that the superoperators $\hat Q^k$
and $\hat P^k$ satisfy the relations which are the
quantum analogs to the relations for
the operators $Q^k$ and $P^k$:\\
1. $\hat Q^k \hat I=\hat x^k$; $\hat P^k \hat I=\hat 0$.\\
2. $\overline{\hat Q^k}=\hat Q^k$; $\overline{\hat P^k}=\hat P^k$. \\
3. $(\hat Q^{k}\hat A)^*=\hat Q^k \hat A^*$;
$(\hat P^{k} \hat A)^*=-\hat P^k \hat A^*$.\\
4. $[\hat Q^k,\hat P^m]=i\delta_{km} \hat I$,
$[\hat Q^k,\hat Q^m]=\hat 0$; $[\hat P^k,\hat P^m]=\hat 0$;
$[\hat I,\hat Q^k]=\hat 0$; $[\hat I ,\hat P^k]=\hat 0$.\\
Superoperator $\hat {\cal L}$ is called formally self-adjoint
$\overline{\hat {\cal L}}=\hat{\cal L}$, if the relation
$<\hat {\cal L}\hat A|\hat B>=<\hat A|\hat {\cal L} \hat B>$ is satisfied.
Here the scalar product $<\hat A|\hat B>$ on the operator space ${\cal M}$
is defined by
\[ <\hat A|\hat B> \equiv Sp[\hat A^* \hat B] \ . \]
An operator space with this scalar product is called Liouville space
\cite{kn1,kn2}.

The Weyl quantization of operators $Q^k$ and $P^k$ is defined by
Weyl quantization of functions $Q^kA(x)$ and $P^k A(x)$
for all $A(x)$.

To quantize the operator $P^k$ we use the relation
\[ \{x^m,A(x)\}=\Psi^{kk'}\partial_{k}x^m \partial_{k'} A(x) \ , \]
where $\partial_k x^m=\delta^m_k$
and $\omega_{km}\Psi^{mk'}=\delta^{k'}_{k}$. Then
\[ \partial_k A(x)=\omega_{km}\{x^m, A(x) \} \ . \]
Quantization of the right-hand side of this formula
leads to the expression
\[ \omega_{km} \frac{1}{i \hbar}[\hat x^m, \hat A(\hat x)] \ . \]
The operator $\hat P^{k}\hat A$  is written in the form
\[ \hat P^k \hat A(\hat x)=-\omega_{km}
\frac{1}{ \hbar}[\hat x^m, \hat A(\hat x) ] \ . \]
As a result, we obtain
\be
\label{Pk}
\hat P^k =-\omega_{km} \frac{1}{ \hbar}[\hat x^m, \ . \ ]
=- \omega_{km} \frac{1}{\hbar}((\hat x^m)^l-(\hat x^m)^r) \ .
\ee
In the usual notations $\hat x^{2j-1}=\hat q^j$
and $\hat x^{2j}=\hat p^j$ this formula can be rewritten
\[ \hat P^{2j-1}=\frac{1}{\hbar}((\hat p^j)^l-(\hat p^j)^r) \ ,\quad
\hat P^{2j}=-\frac{1}{\hbar}((\hat q^j)^l-(\hat q^j)^r) \ ,  \]
where the left and right superoperators are defined by
\[ (\hat q^j)^l \hat A= \hat q^j \hat A \ , \quad
(\hat p^j)^l \hat A= \hat p^j \hat A \ , \quad
(\hat q^j)^r \hat A= \hat A \hat q^j \ , \quad
(\hat p^j)^r \hat A= \hat A \hat p^j \ . \]
for all $\hat A \in \hat {\cal M}$.

Let us quantize the operator $Q^k$.
It is known \cite{Ber1} that the Weyl quantization of the expression
$x^k \circ A(x)$ leads to $\hat x^k \circ \hat A(\hat x)$.
Therefore superoperator $\hat Q^k$ has the form
\be
\label{Qk}
\hat Q^k=[\hat x^k, \ . \ ]_{+}= \frac{1}{2} ((\hat x^k)^l+(\hat x^k)^r) \ ,
\ee
i.e. $\hat Q^k \hat A=\hat x^k \circ \hat A$.

It can be verified that for the superoperators $\hat Q^k$ and $\hat P^k$
defined by (\ref{Pk}) and (\ref{Qk}) the commutation relations
\be  \label{QP-cr}
[\hat Q^k,\hat P^m]=i\delta_{km} \hat I \ , \quad
[\hat Q^k,\hat Q^m]=0 \ ,  \quad [\hat P^k,\hat P^m]=0  \ , \ee
are satisfied. To check these relations, we must express
the superoperators $\hat Q^k$ and $\hat P^k$ via superoperators
$(\hat x^k)^l$ and $(\hat x^k)^r$
and use the commutation relations
\be \label{ff1}
[(\hat x^k)^l,(\hat x^m)^l]=i \hbar \Psi^{km} \hat I \ ,
\quad [(\hat x^k)^r,(\hat x^m)^r]=-i \hbar \Psi^{km} \hat I \ ,
\quad [(\hat x^k)^l,(\hat x^m)^r]=0 \ ,
\ee
which follow from canonical commutation relations (\ref{ccr}).
Using relations (\ref{Pk}) and (\ref{Qk}), the other relations
for the superoperators $\hat P^{k}$ and $\hat Q^{k}$
can be easy to verified.

Weyl quantization of the operators (\ref{V1}) leads to the superoperators
\be
\label{HV1}
\hat V(a,b,\hat Q,\hat P)=exp\{ i(a \hat Q+b \hat P)\} \ .
\ee
The relations (\ref{QP-cr}) for $\hat Q^k$ and $\hat P^k$
leads to the relations:
\be
\label{V21}
\hat V(a_1,b_1,\hat Q,\hat P) \hat V(a_2,b_2,\hat Q,\hat P)=
\hat V(a_1+a_2,b_1+b_2,\hat Q,\hat P)
exp\{ -\frac{i}{2}[a_1b_2-a_2b_1] \} \ ,
\ee
\be
\label{V31}
\hat V^{\dagger}(a,b,\hat Q,\hat P)=\hat V(-a,-b,\hat Q,\hat P) \ ,
\quad \hat V^{\dagger}(a,b,\hat Q,\hat P)
\hat V(a,b,\hat Q,\hat P)=\hat I \ .
\ee
It allows to define superoperator Weyl algebra which is involute normed
algebra. The involution corresponds to conjugation (\ref{V31}).
A superoperator norm \cite{kn1,kn2} defines the norm of the algebra.
The algebra $\hat {\cal A}(\hat {\cal M})$ of bounded dynamical
superoperators can be defined as $C^{*}$-algebra.
It contains all superoperators $\hat V(a,b,\hat Q,\hat P)$ and it is closed
for linear combinations of $\hat V(a,b,\hat Q,\hat P)$ in superoperator norm
topology.

The operator $Q^k$ realizes multiplication by $x^k$ in the Jordan
algebra $J({\cal M})$. Therefore the superoperator $\hat Q^k$
realizes multiplication by the operator $\hat x^k$ on the Jordan
algebra $\hat J(\hat {\cal M})$. The operator $P^k$ represents
multiplication by $x^k$ on the Lie algebra $L({\cal M})$ and
superoperator $\hat P^k$ is multiplication by $\hat x^k$ on the
Lie algebra $\hat L(\hat {\cal M})$. Therefore the Weyl
quantization of operators $Q^k$ and $P^k$: is realized as a map of
algebras of multiplication operators which act on the Jordan and
Lie algebras. Note that operators $Q^{k}$, $P^{k}$ and
superoperators $\hat Q^{k}$, $\hat P^{k}$ can be defined by
\[  Q^k A(x)=g_{Jord}(x^k,A(x)) \ , \quad
P^k A=g_{Lie}(x^k,A(x)) \ , \]
\[ \hat Q^k \hat A =  \hat g_{Jord}(\hat x^k, \hat A) \ , \quad
\hat P^k \hat A = \hat g_{Lie}(\hat x^k, \hat A) \ . \]


\section{Quantization of Operator Function}


Let us consider the dynamical operator ${\cal L}={\cal L}(Q,P)$
as a function of the basis operators $Q^k$ and $P^k$.
Generalized quantization can defined as a map from
dynamical operator space $A({\cal M})$ to dynamical superoperator
space $\hat A(\hat {\cal M})$.
The Weyl quantization of the symmetric form of operator
\[ {\cal L}(Q,P)=\frac{1}{(2\pi)^{2n}} \int \tilde L(a,b) e^{i(aQ+bP)} da db
\ , \quad Q^k=x^k \ , \ \ P^k=-i \partial_k \ , \]
leads to the corresponding superoperator
\be
\label{f11}
\hat {\cal L}(\hat Q,\hat P)=\frac{1}{(2\pi)^{2n}}
\int \tilde L(a,b) e^{i(a \hat Q+b \hat P)} da db \ ,
\ee
\[ \hat Q^k=\frac{1}{2}((\hat x^k)^l+(\hat x^k)^r) \ , \ \
\hat P^k=-\frac{1}{\hbar}\omega_{km}((\hat x^m)^l-(\hat x^m)^r) \ . \]

If the function $ \tilde L(a,b)$ is connected with Fourier image
$\tilde A(a)$ of the function $A(x)$ by the relation
\[ \tilde L(a,b)=(2 \pi)^n \delta (b) \tilde A(a) \ , \]
then the formula (\ref{f11}) defines the canonical quantization of
the function $A(x)={\cal L}(Q,P)1$ by the relation $\hat A(\hat
x)=\hat {\cal L}(\hat Q,\hat P) \hat I$. Here we use $\hat Q^k
\hat I=\hat x^k$ and $exp\{ia \hat Q\} \hat I= exp\{ia \hat x\}$.
Therefore canonical quantization is a specific case of suggested
quantization procedure. If we use $QP$-form of the dynamical
operator ${\cal L}={\cal L}(Q,P)$, then the Weyl quantization of
\[ {\cal L}(Q,P)=\frac{1}{(2\pi)^{2n}} \int L(a,b)e^{iaQ}e^{ibP} da db \ , \]
leads to the superoperator
\be
\label{f112}
\hat {\cal L}(\hat Q,\hat P)=\frac{1}{(2\pi)^{2n}}
\int L(a,b) e^{ia \hat Q} e^{ib \hat P} da db \ ,
\ee
Superoperators $\hat Q^k$ and $\hat P^k$ can be represented by
$(\hat x^k)^l$ and $(\hat x^k)^r$.
Therefore the formula (\ref{f11}) can be written in the form
\be
\label{f12}
\hat {\cal L}( \hat x^l,  \hat x^r)=\frac{1}{(2\pi)^{2n}} \int
\tilde L(a',b') W^l(a', \hat x)W^r(b', \hat x) da' db' \ .
\ee
Here $\hat W^l(a, \hat x)$ and $\hat W^l(a, \hat x)$ are left and
right superoperators corresponding to the Weyl operator (\ref{f2}).
These superoperators can be defined by
\[ \hat W^{l}(a,\hat x)=\hat W(a,\hat x^{l}) \ , \quad
 \hat W^{r}(a,\hat x)=\hat W(a,\hat x^{r}) \ . \]
If the function $\tilde L(a',b')$ of the superoperator (\ref{f12})
has the form
\[ \tilde L(a',b')=\frac{i}{\hbar}(H(a')\delta(b')-H(b')\delta(a')) \ , \]
then quantum system is Hamiltonian system and Heisenberg equation
has the usual form:
\[ \frac{d}{dt} \hat A_t= \frac{i}{\hbar}[\hat H(\hat x), \hat A_t] \ . \]
Here $\hat H(\hat x)$ is Hamilton operator defined by
\[ \hat H(\hat x)=\frac{1}{(2\pi \hbar)^{n}} \int
H(a') \hat W(a', \hat x) da' \ . \]


\section{Explicit Formulas of Quantization}


Let us derive a relation which represents the superoperator
$\hat{\cal L}(\hat Q, \hat P)$ by operator ${\cal L}(Q,P)$.
We would like to find the analog of the relation (\ref{f4'})
between an operator $\hat A(\hat x)$ and a function $A(x)$.
In order to find the relation, it will be convenient to represent
the Fourier image of a function $\tilde L(a,b)$
by the operator ${\cal L}(Q,P)$.

Let us find the symbol of a dynamical operator ${\cal L}(Q,P)$
by this operator.
To simplify formulas, we introduce new notations.
Let $X^s$, where $s=1,...,4n$, denote the operators $Q^k$ and $P^k$,
where $k=1,...,2n$, that is $Q^k=X^k$ and $P^k=X^{k+2n}$, or
\[ X^{2j-1}=q^j \ , \quad X^{2j}=p^j \ , \quad
X^{2j-1+2n}=-i\frac{\partial}{\partial q^j} \ , \quad
X^{2j+2n}=-i \frac{\partial}{\partial p^j} \ , \]
where $j=1,...,n$.
Let us denote the parameters $a^k$ and $b^k$, where $k=1,...,2n$, by
$z^s$, where $s=1,...,4n$.
Then the formula (\ref{f9}) can be rewritten by
\[ {\cal L}(X)=\frac{1}{(2\pi)^{2n}} \int \tilde L(z) e^{izX} dz \ . \]
The formula (\ref{f11}) for the superoperator $\hat {\cal L}$
is written in the form
\[ \hat {\cal L}(\hat X)=\frac{1}{(2\pi)^{2n}}
\int \tilde L(z) e^{iz \hat X} dz \ . \]
Let us consider the inverse Fourier transformation for the function $L(z)$:
\[ {\cal L}(\alpha)=\frac{1}{(2\pi)^{2n}}
\int \tilde L(z) e^{iz \alpha} dz \ , \quad  \tilde L(z)=\frac{1}{(2\pi)^{2n}}
\int {\cal L}(\alpha) e^{-iz \alpha} d \alpha \ . \]
Here $\tilde L(z)$ is the Fourier image of the function ${\cal L}(\alpha)$.
Then operator ${\cal L}(X)$ can be written by the Weyl symbol
${\cal L}(\alpha)$ in the form
\be
\label{lx}
{\cal L}(X)=\frac{1}{(2\pi)^{4n}}
\int {\cal L} (\alpha) e^{iz (X-\alpha)} dz d \alpha \ .
\ee
This formula is an analog of the formula (\ref{f4'}).
The generalized quantization of operators $X^{s}$
leads to the superoperators $\hat X^{s}$. Therefore Weyl quantization of the
operator (\ref{lx}) leads to the superoperator
$\hat {\cal L}(\hat X)$ which can be written
\be
\label{ff2}
\hat {\cal L}(\hat X)=\frac{1}{(2\pi)^{4n}}
\int {\cal L}(\alpha) e^{iz (\hat X-\alpha)} dz d \alpha \ .
\ee

Let ${\cal L}_m(\alpha)$, where $m=1$ or $2$, are the Weyl symbols of the
operator ${\cal L}_m(X)$. Then
\be
\label{Lm}
{\cal L}_m(X)=\frac{1}{(2\pi)^{4n}}
\int {\cal L}_m(\alpha) e^{iz (X-\alpha)} dz d \alpha \ .
\ee

Let us consider the trace $Sp[{\cal L}_1 (X) {\cal L}_2(X)]$ of the
product of the operators ${\cal L}_m(X)$ and find
the Weyl symbol ${\cal L}_2(\alpha)$ of the operator ${\cal L}_2(X)$.
As follows from (\ref{ff2}), we have
\[ Sp[{\cal L}_2(X){\cal L}_1(X)]=\frac{1}{(2\pi)^{4n}} \int
{\cal L}_1(\alpha) Sp[ {\cal L}_2(X) e^{iz (X-\alpha)}] dz d \alpha \ . \]
This relation can be compared with the well known formula
for the trace of operator product

\[ Sp[{\cal L}_2(X){\cal L}_1(X)]=\frac{1}{(2\pi)^{4n}}
\int {\cal L}_1(\alpha) {\cal L}_2(\alpha) d \alpha \ . \]
We obtain the relation for the symbol ${\cal L}_2(\alpha)$ of the operator
${\cal L}_2(X)$
\[ {\cal L}_2(\alpha)=\int Sp[ {\cal L}_2(X) e^{iz (X-\alpha)}] dz \ . \]
The formula for the Weyl symbol ${\cal L}(\alpha)$ of the operator
${\cal L}(X)$ has the form
\[ {\cal L}(\alpha)=\int e^{-iz'\alpha} Sp[ {\cal L}(X) e^{iz' X}] dz' \ . \]
Next we substitute this symbol into the relation (\ref{ff2}).
As the result we obtain
\be
\label{ff3}
\hat {\cal L}(\hat X)=\frac{1}{(2\pi)^{4n}}
\int e^{-i\alpha(z+z')} e^{iz\hat X}
Sp[{\cal L}(X) e^{iz' X}] dz d \alpha dz' \ .
\ee

In the new notations, the operators $V(a,b,Q,P)$ are written
as $V(z,X)=exp\{izX\}$.
If we use the superoperator $\hat V(z,\hat X)=exp\{i z \hat X\}$,
then the formula (\ref{ff3}) can be rewritten in the form
\[ \hat {\cal L}(\hat X)=\frac{1}{(2\pi)^{4n}} \int
e^{-i\alpha(z+z')} \hat V(z,\hat X) Sp[{\cal L}(X) V(z', X)]
dz d \alpha dz' \ . \]


\section{Harmonic oscillator with friction}


Let us consider $n$-dimensional linear oscillator with friction
$F^{j}_{fric}=-(\gamma/m)p^{j}$. The time evolution equation
for this oscillator has the form
\[ \frac{d}{dt}q^j=\frac{1}{m}p^j \ , \quad
\frac{d}{dt} p^j=-(m\omega^2 q^j+\frac{\gamma}{m} p^j) \ . \]
The dynamical equation for the classical observable $A_{t}(q,p)$ is written
\[ \frac{d}{dt}A_t(q,p)={\cal L}(q,p,\partial_q, \partial_p) A_t(q,p) \ . \]
Differentiation of the function $A_{t}(q,p))$ gives
\[ \frac{d A_t (q,p)}{dt} =\frac{\partial A_t(q,p)}{\partial q^j}
\frac{dq^j}{dt} +\frac{\partial A_t(q,p)}{\partial p^j} \frac{dp^j}{dt}= \]
\[ =\frac{1}{m} p^j\frac{\partial A_t (q,p)}{\partial q^j}-
(m\omega^2 q^j+\frac{\gamma}{m} p^j)
\frac{\partial A_t (q,p)}{\partial p^j} \ . \]
The dynamical operator ${\cal L}(q,p,\partial_q, \partial_p)$ is
\[ {\cal L}(q,p,\partial_q, \partial_p)=
\frac{1}{m}p^j\frac{\partial}{\partial q^j}-
(m\omega^2 q^j+\frac{\gamma}{m} p^j)\frac{\partial}{\partial p^j} \ . \]
This operator is written in the $QP$-form
\be \label{L1}
{\cal L}(Q,P)=\frac{i}{m}Q^{2j}P^{2j-1}-
i(m\omega^2 Q^{2j-1}+\frac{\gamma}{m} Q^{2j}) P^{2j} \ .
\ee
Weyl quantization of the operator
${\cal L}={\cal L}(Q,P)$ leads to superoperator
\be \label{LQP}
\hat {\cal L}(\hat Q, \hat P)=\frac{i}{m}\hat Q^{2j} \hat P^{2j-1}-
i(m\omega^2 \hat Q^{2j-1}+\frac{\gamma}{m} \hat Q^{2j}) \hat P^{2j} \ .
\ee
This superoperator can be written as
\[ \hat {\cal L}(\hat Q, \hat P)=
\frac{i}{2 m \hbar}[(\hat p^2+m^2\omega^2 \hat q^2)^l-
(\hat p^2+m^2\omega^2 \hat q^2)^r]+\frac{i \gamma}{2 m \hbar}[(\hat p
\hat q )^l-(\hat p \hat q)^r+ \hat q^l \hat p^r-\hat p^l \hat q^r] \ . \]
As the result we have generalized Heisenberg equation
\[ \frac{d}{dt} \hat A_t=\frac{i}{\hbar}[\hat H, \hat A_{t}]+
\frac{i \gamma}{m\hbar}[\hat p^j,[\hat q^j, \hat A_{t}]]_{+} \ , \]
where
\[ \hat H=\frac{1}{2m}(\hat p^2+m^2\omega^2 \hat q^2) \ . \]


\section{Fokker-Planck-Type System}


Let us consider Liouville operator $\Lambda$, which acts on the
normed distribution density function $\rho(q,p,t)$ and has the form
of second order differential operator
\be
\label{liu-li}
\Lambda=d_{qq} \frac{\partial^{2}}{\partial q^{2}}+
2 d_{qp} \frac{\partial^{2}}{\partial q \partial p}+
d_{pp} \frac{\partial^{2}}{\partial p^{2}}+
+c_{qq} q \frac{\partial}{\partial q}+
c_{qp} q \frac{\partial}{\partial p}+
c_{pq} p \frac{\partial}{\partial q}+
c_{pp} p \frac{\partial}{\partial p}+h \ .
\ee
Liouville equation
\[ \frac{d \rho(q,p,t)}{d t}= \Lambda \rho(q,p,t) \]
with operator (\ref{liu-li})
is Fokker-Planck-type equation. Weyl quantization of the Liouville
operator (\ref{liu-li}) leads to completely dissipative superoperator
$\hat \Lambda$ considered in \cite{Lind1,Sand,kn1}.
As the result we have the quantum Markovian master equation
for a matrix density operator $\hat \rho_{t}$.
If $h=-2(c_{pp}+c_{qq})$, then this equation has  the form
\[ \frac{d\hat \rho_t}{dt}=-\frac{i}{\hbar}[\hat H, \hat \rho_t]+
\frac{i(\lambda-\mu)}{\hbar}[\hat p,\hat q \circ \hat \rho_t]-
\frac{i(\lambda+\mu)}{\hbar}[\hat q,\hat p \circ \hat \rho_t]- \]
\be \label{8.16}
-\frac{d_{pp}}{{\hbar}^2}[\hat q,[\hat q,\hat \rho_t]]
-\frac{d_{qq}}{{\hbar}^2}[\hat p,[\hat p, \hat \rho_t]]
+\frac{2d_{pq}}{{\hbar}^2}[\hat p,[ \hat q,\hat \rho_t]] \ .
\ee
Here $\hat H$ is Hamilton operator which has the form
\be
\label{8.14}
\hat H=\frac{1}{2m}\hat p^2+\frac{m\omega^2}{2} \hat q^{2} \ , \ee
where
\[ m=-\frac{1}{c_{pq}} \ , \quad \omega^{2}=-c_{qp}c_{pq} \ , \quad
\lambda=\frac{1}{2}(c_{pp}+c_{qq}) \ , \quad
\mu=\frac{1}{2}(c_{pp}-c_{qq}) \ . \]
Here $d_{pp}$, $d_{qq}$, $d_{pq}$ are quantum diffusion coefficients and
$\lambda$ is a friction constant.


\section{Conclusions}


In this paper we suggest a generalization of Weyl quantization
called dynamical quantization. It allows to derive
dynamical superoperator from dynamical operator.
The basis formulas which define the suggested quantization are
(\ref{f11}) and (\ref{ff3}).
Quantization of a general dynamical operator for non-Hamiltonian system
is described by suggested dynamical quantization. The suggested Weyl
quantization scheme  allows to derive quantum analogs for the
classical non-Hamiltonian and dissipative systems.

The dynamical quantization (\ref{f11}) and (\ref{ff3}) map the operator
${\cal L}(q,p,\partial_{q},\partial_{p})$ which
acts by the functions $A(q,p)$ to the superoperator $\hat {\cal L}$,
which acts on the elements of operator space.
If the operator ${\cal L}$ is an operator ${\cal L}(q,p)$ of
multiplication on the function $A(q,p)={\cal L}(q,p)1$, then
formula (\ref{ff3}) defines the usual Weyl quantization of the function
$A(q,p)$ by the relation $\hat A=\hat {\cal L} \hat I$.
Therefore the usual Weyl quantization procedure is a specific case of
suggested Weyl dynamical quantization.

This work was partially supported by the RFBR grant No. 00-02-17679.

\vskip 11 mm



\end{document}